\documentclass{amsart}

\usepackage{amssymb, amsmath}

\usepackage{euscript}

\begin{document}

\def\We{\mathrm{We}}

\def\wh{\widehat}
\def\wt{\widetilde}
\def\ov{\overline}

\def\Sp{\mathrm  {Sp}}
\def\GL{\mathrm  {GL}}
\def\Heis{\mathrm  {Heis}}
\def\SL{\mathrm  {SL}}

\def\C{\mathbb C}
\def\R{\mathbb R}
\def\Z{\mathbb Z}
\def\Q{\mathbb Q}
\def\A{\mathbb A}
\def\T{\mathbb T}
\def\U{\mathbb U}
\def\H{\mathbb H}
\def\K{\mathbb K}

\def\QED{\hfill $\square$}

\def\B{\mathrm B}

\def\epsilon{\varepsilon}

\def\cA{\mathcal A}
\def\cZ{\mathcal Z}
\def\cV{\mathcal V}
\def\cL{\mathcal L}
\def\cD{\mathcal D}
\def\cP{\mathcal P}
\def\cM{\mathcal M}
\def\cS{\mathcal S}
\def\cB{\mathcal B}
\def\cH{\mathcal H}
\def\cF{\mathcal F}
\def\cI{\mathcal I}

\def\lr{\leftrightarrow}
\def\lra{\longrightarrow}
\def\llr{\longleftrightarrow}
\def\LR{\Longleftrightarrow}
\def\RA{\Longrightarrow}
\def\LA{\Longleftarrow}
\def\Lr{\Leftrightarrow}
\def\Ra{\Rightarrow}
\def\La{\Leftarrow}

\renewcommand{\Re}{\mathop{\rm Re}\nolimits}

\renewcommand{\Im}{\mathop{\rm Im}\nolimits}

\newcounter{sec} \renewcommand{\theequation}{\arabic{sec}.\arabic{equation}}
\newcounter{punct}[sec]

\newcounter{fact} \def\fact{\addtocounter{fact}{1}{\scc \arabic{fact}}}

\renewcommand{\thepunct}{\thesec.\arabic{punct}}
\def\punct{\refstepcounter{punct}{\arabic{sec}.\arabic{punct}.  }}

\stepcounter{sec}

\def\SS{\smallskip}

\newtheorem{theorem}{Theorem}[sec]
\newtheorem{proposition}[theorem]{Proposition}
\newtheorem{prop}[theorem]{Proposition}
\newtheorem{lemma}[theorem]{Lemma}
\newtheorem{cor}[theorem]{Corollary}
\newtheorem{corollary}[theorem]{Corollary}
\newtheorem{observation}[theorem]{Observation} 

\begin{center}
\bf\Large
On adelic model of boson Fock space

\medskip

\sc\large
Neretin Yu.A.%
\footnote{supported by grant NWO--047.017.015}  
\end{center}

{\small We construct a canonical embedding
of the Schwartz space on $\R^n$ to the space of 
distributions on the adelic product of all the $p$-adic
numbers. This map is equivariant 
with respect to the action of the symplectic
group $\Sp(2n, \Q)$ over rational numbers and with respect
to the  action of rational Heisenberg group.}

\bigskip

These notes contain two elements. First, we give a 
funny realization of a space of complex functions of a real variable
as a space of functions of $p$-adic variable. Secondly, 
we try to clarify classical contstruction of modular forms
through $\theta$-functions and Howe duality.


\section{Introduction}

{\bf\punct Fields and rings.} 
Below $\Q$ denotes the rational numbers,
$\Z$ is the ring of integers, $\Q_p$ is the field of $p$-adic
numbers, $\Z_p\subset \Q_p$
 is the ring of $p$-adic integers.
We denote the norm on $\Q_p$ by $|\cdot|$.


{\bf \punct Adeles,}
(see 
\cite{GGP}, \cite{Wei}, \cite{PR}).
  An {\it adele} is a sequence
\begin{equation}
(a_\infty, a_2,a_3,a_5,a_7,a_{11},\dots),
\label{R-A}
\end{equation}
where $a_\infty\in\R$, $a_p\in \Q_p$ ($p$ is a prime)
 and $|a_p|=1$ for all $p$ 
except a finite a number of primes.

Our main object is the ring 
$$
\A\subset \Q_2\times \Q_3 \times \Q_5\times \dots 
$$
 consisting
of the sequences
$$
a=(a_2,a_3,a_5,a_7,a_{11},\dots)
$$
satisfying the same conditions. The addition and multiplication
in $\A$ are defined coordinate-wise.
Below the term {\it "adeles"} means the ring $\A$.
The space of sequences of the form (\ref{R-A}) we denote
by $\R\times\A$.

\SS


{\bf\punct Convergence in $\A$.}
A sequence $a^{(j)}$ in $\A$  converges to $a\in A$
iff 

a) There is a finite set $S$ of primes such that
$|a_p^{(j)}|=1$ for all $p\notin S$ for all $j$.

b) For each $p$, the sequence $a^{(j)}_p$ converges in $\Q_p$.

\SS

The image of the diagonal embedding
$\Q\to\A$
$$r\mapsto(r,r,r,\dots)$$
is dense in $\A$.


\SS

{\bf\punct Integration.} We define the Lesbegue measure
$da$
on the ring $\A$ by two assumptions: 

-- the measure on $\prod_p \Z_p$ is the product-measure

-- the measure is translation-invariant.

This measure is $\sigma$-finite. We define the space $L^2(\A^n)$
in the usual way. The Bruhat space $\cB(\A^n)$ defined below
is dense in $L^2(\A^n)$.

\SS


{\bf\punct Adelic exponents.}
For an adele $a\in\A$, we define its exponent $\exp(2\pi i a)\in\C$ by
$$
\exp(2\pi i a) =\prod_p \exp(2\pi i a_p).
$$   
all the factors are roots of unity, only finite
number of factors is $\ne 1$.


{\bf \punct Lattices.}
A {\it lattice} $L$ in a $\Q$-linear space $\Q^n$
is an arbitrary additive subgroup isomorphic $\Z^n$.
Equivalently, a lattice is  a group
$L\subset \Q^n$ having a form $\bigoplus \Z f_j$, where $f_j$
is a basis in  $\Q^n$.
A {\it dual lattice} $L^\lozenge$ consists of $y\in\Q^n$, such that
$\sum x_j y_j\in\Z$ for all the $x\in L$.   

A {\it lattice} in the $p$-adic 
linear space $\Q_p^n$ is a set of 
the form $\bigoplus \Z_p f_j$, where $f_j$ is a basis
in $\Q_p^n$. The {\it standard lattice}
is the set $\Z_p^n$.

A {\it lattice} in the adelic module $\A^n$  is a set
of a form
$\bigoplus_p L_p,$
where $L_p\subset \Q_p^n$ are lattices, and 
$L_p$ are the standard lattices for all  
$p$ except a finite set.

For a lattice
$L\subset \Q^n$, consider its closure $\ov{\ov L}\subset \A^n$.
It is a lattice, and moreover the map $L\mapsto \ov{\ov L}$
is a bijection of the set of  all the lattices in $\Q^n$ and the 
set 
of all the lattices in $\A^n$.


{\bf\punct Bruhat test functions and distributions
on $\A^n$.} A {\it test function} $f$ on $\Q_p^n$ or on $\A^n$  is a
compactly supported locally constant complex-valued function.
The Bruhat space $\cB(\Q_p^n)$ (resp. $\cB(\A^n)$)
is the space of all the test functions.

A {\it distribution} is a linear functional
on $\cB(\Q_p^n)$ (resp. $\cB(\A^n)$). We denote the space of all the 
distributions by $\cB'(\Q_p^n)$ (resp.  $\cB'(\A^n)$).


{\bf \punct The second description of the spaces $\cB$.}
 Let $S$ be a subset in $\Q_p^n$ or $\A^n$.
Denote by $\cI_S$ the indicator function
of $S$, i.e. 
$$\cI_S(x)= 
\begin{cases}
\text{1,  if $x\in S$}
\\
\text{0,  if $x\in S$}
.\end{cases}
$$
 
For a lattice $L$ and a vector $a$, the function 
$
\cI_{L+a}
\label{test}
$
 is a test function. Each test function is
a linear combination of functions of this type.
  

{\bf\punct Third description of the spaces $\cB$.}
Consider the space $\Q_p^n$ or $\A^n$.
Let $K\subset L$ be lattices.
 Denote by $\cB(L|K)$ the space

a) $f=0$ outside $L$.

b) $f$ is $K$-invariant.

The dimension of this space is the order of the quotient group
$L/K$, in particular the dimension is finite.

Then
$$
\cB(\Q_p^n)=\bigcup_{K\subset L\subset\Q_p^n}\cB(L|K; \Q_p),\qquad
\cB(\A^n)=\bigcup_{K\subset L\subset\A^n}\cB(L|K; \A)
.$$


{\bf \punct The space $\cM(\Q^n)$.} We repeat literally
the previous definition. For two lattices $K\subset L\subset \Q^n$,
denote by 
$\cM(L|K)$ the space of $K$-invariant functions on $\Q^n$
supported by $L$. We assume
$$
\cM(\Q^n)= \bigcup_{K\subset L\subset\Q^n}\cM(L|K).
$$

The space $\cM(\Q^n)$ is generated by the indicator functions
$\cI_{L+a}$ of shifted lattices.


\SS

{\bf\punct The bijection $\cM(\Q^n)\lr\cB(\A^n)$.}

\begin{proposition}
a) Each function $f\in\cM(\Q^n)$ admits a unique continuous
extension $\ov{\ov f}$ to a function on $\A^n$

b) The map $f\mapsto \ov{\ov f}$ is a bijection $\cM(\Q^n)\to\cB(\A^n)$.
\end{proposition}

The statement is trivial. More constructive variant of this is statements
is
$$
\ov{\ov{\cI_{L+a}}}=\cI_{\ov{\ov L}+a}
.$$


{\bf\punct Space  $\cP(\R^n)$ of Poisson distributions.}
 Denote by $\cS(\R^n)$ the {\it Schwartz
space} on $\R^n$, i.e. the space of smooth functions $f$ on $\R^n$ 
satisfying the condition:
 for each $\alpha_1$,\dots, $\alpha_n$, and each $N$
$$
\lim_{x\to\infty} \left(\sum x_j^2\right)^N \frac{\partial^{\alpha_1}}{\partial^{\alpha_1} x_1}
\dots \frac{\partial^{\alpha_n}}{\partial^{\alpha_n} x_n}
f(x)=0
.$$

By $\cS'(\R^n)$ denote the space dual to $\cS(\R^n)$, i.e.,
the space of all {\it tempered distributions} on $\R^n$.

Now we intend to define a certain 
 subspace $\cP(\R^n)\subset \cS'(\R^n)$.
This space is spanned by functions
$$
\sum_{k_1,\dots,k_n\in \Z}
\delta\bigl(x - \sqrt{2\pi}(b+\sum_j k_j a_j )\bigr)   
,$$
where $a_1$, \dots, $a_n\in \Q^n$
 are linear independent, $b\in\Q^n$.

\begin{lemma}
 A  countable sum $\psi$ of $\delta$-functions
is an element of $\cP(\R^n)$ iff there are two lattices
$K\subset L\subset \Q^n$ such that $\psi$ is supported by
$\sqrt{2\pi}L$ and $\psi$ is $\sqrt{2\pi}K$-invariant.
\end{lemma}

We denote by $\cP(L|K)\subset \cP(\R^n)$ the space of all
the distributions satisfying this Lemma.

\SS


{\bf \punct Canonical bijection $\cM(\Q^n)\lr \cP(\R^n)$.} 
Define a canonical bijective map $I_\R:\cM(\Q^n)\to \cP(\R^n)$. 
Let $f\in \cM(\Q^n)$, let $M\subset L\subset \Q^n$ 
be corresponding lattices. We define the 
distribution $I_\R f\in \cP(\R^n)$ as 
$$
I_\R f(x)=\sum_{\xi\in L} f(\xi)\delta(x-\sqrt{2\pi}\xi)
.$$
We obtain the bijection
$\cM(\Q^n)\to \cP(\R^n)$. Also, for each
rational  lattices $K\subset L$, we have a bijection
$$
\cM(L|K)\llr\cP(L|K)
.$$


{\bf \punct Observation.} Thus we have the canonical
bijection 
 \begin{equation}
J_{\R\A}:
\Bigl\{\text{space $\cP(\R^n)$}\Bigr\}
 \llr 
\Bigl\{\text{adelic space $\cB(\A^n)$}
\Bigr\}
\label{main}
.\end{equation}
In particular, we have canonical embeddings
\begin{align*}
\cS(\R^n)\to\cB'(\A^n)
,
\\
\cB(\A^n)\to\cS'(\R^n)
.\end{align*}

{\bf\punct The image of the  Schwartz space in $\cB'(\A^n)$.}

\begin{proposition}
For $f\in\cS(\R^n)$, the corresponding 
element $F\in\cB'(\A^n)$
is 
\begin{equation}
F(a)=\sum_{\xi\in \Q^n} f(\xi)\delta_\A(a-\xi)
\label{adelic-delta}
\end{equation}  
where $\delta_\A$ is the adelic delta-function.
\end{proposition}

{\sc Proof.}
Let $L\subset \Q$ be a lattice, $b\in\Q^n$.
 The value of the adelic distribution $F\in \cB'(\A^n)$ on
the adelic test function $\cI(\ov{\ov L}+b)$
is 
$$
\sum_{\xi\in\bigl[\Q^n\cap (\ov{\ov L}+b)\bigr]} f(\xi)
=\sum_{\xi\in (L+b)} f(\xi)
$$
The last expression is the value of the Poisson distribution
$I_\R \cI_{L+a}$ on the Schwartz function $f$.\QED

\SS


{\bf \punct Result of the paper.}
 The space $\cS(\R^n)$
is equipped with the canonical action of the real
Heisenberg group%
\footnote{All the definitions are given below.}
 $\Heis_n(\R)$  and the real symplectic
group $\Sp(2n,\R)$ (in this sense,
 $\cS(\R^n)$ is a {\it bosonic Fock space}
mentioned in the title).

The space $\cB(\A^n)$ is equipped with the canonical action
of the adelic Heisenberg group $\Heis_n(\A)$ and the adelic
symplectic group $\Sp(2n,\A)$.

There are canonical embeddings 
\begin{align*}
\Heis_n(\Q)\to \Heis_n(\R),\qquad 
   \Heis_n(\Q)\to \Heis_n(\A),\\
\Sp(2n,\Q)\to\Sp(2n,\R),\qquad
\Sp(2n,\Q)\to\Sp(2n,\A);
\end{align*}
in all the cases the images are dense.

\begin{theorem}
\label{th:main}
a) The map $J_{\R\A}$ commutes with the action of $\Heis_n(\Q)$.

b) The map $J_{\R\A}$ commutes with the action of $\Sp(2n,\Q)$.
\end{theorem}

\begin{corollary} 
For $f\in\cS(\R)$ denote by $\widehat f$ its Fourier transform.
Then the adelic Fourier transform of the distribution
(\ref{adelic-delta}) is 
$$
\mathrm{const}\cdot
\sum_{\xi\in \Q^n} \widehat f(\xi)\delta_\A(a-\xi)
$$
\end{corollary}

\begin{theorem}
\label{th:modular}
For each $f\in \cP(\R^n)$, there is a congruence
subgroup in $\Sp(2n,\Z)$ that fixes $f$.
\end{theorem}

{\bf \punct Another description of the operator $J_{\R\A}$.}
Consider the space $\R^n\times \A^n$ (in fact, it is the adelic space
in the usual sense). Consider the tensor product
$\cS(\R^n)\otimes \cB(\A^n)$, and consider the linear functional
(Poisson--Weil distribution) on 
this space given by 
$$
K(x,\xi)=\sum_{\xi\in\Q^n}
\delta_{\R^n}(x+\xi)\delta_{\A^n}(a-\xi)
$$
Our operator $\cS(\R^n)\to \cB'(\A^n)$ is the pairing 
$$
f(x)\mapsto F(a)=\bigl\{ K(x,a),f(x)\bigr\}
$$


\section{Rational Heisenberg group}

\addtocounter{sec}{1}
\setcounter{equation}{0}
\setcounter{punct}{0}
\setcounter{theorem}{0}

%

{\bf\punct Heisenberg group.}
By $\Heis_n$ we denote the group of 
$(1+n+1)\times (1+n+1)$-matrices
\begin{equation}
R(v_+,v_-,\alpha) =
\begin{pmatrix} 
1&v_+&  \alpha+\frac12 v_+v_-^t\\
0&1&v_-^t\\
0&0&1
\end{pmatrix}
\label{heis-matrix}
.\end{equation}
Here $v_+$, $v_-$ are matrices-rows, $v_-^t$ is a matrix-column,
the sign $^t$ is the transposition. We have
\begin{equation}
R(v_+,v_-,\alpha) R(w_+,w_-,\beta)= 
R(v_+ +w_+,v_- + w_-,\alpha+\beta + \frac12(v_+ w_-^t-w_+ v_-^t) )
\label{heis-product}
\end{equation}

We consider 4 Heisenberg groups, $\Heis_n(\Q)$, 
$\Heis_n(\R)$, $\Heis(\Q_p)$,
 $\Heis_n(\A)$, this means that matrix elements
of (\ref{heis-matrix}) are elements of $\Q$, $\R$, $\Q_p$, $\A$.

 The group $\Heis_n(\Q)$
is a dense subgroup in $\Heis_n(\R)$, $\Heis_n(\Q_p)$ $\Heis_n(\A)$. 


{\bf \punct The standard representations of Heisenberg groups.}
These representations are given
by almost the same formulae for the rings $\R$, $\Q$, $\Q_p$, $\A$, but
these formulae differs by position of factors $2\pi$.

\SS

{\it Real case.}
The group $\Heis_n(\R)$ acts in the Schwartz space
$\cS(\R^n)$
on $\R^n$ by  the transformations
$$
T_\R(v_+,v_-,\alpha)f(x)=
f(\sqrt{2\pi}(x+v_+))
\exp\bigl\{\sqrt{2\pi}i xv_-^t+ 2\pi i (\alpha+ \frac12 w_+ v_-^t)\bigr\} 
.$$ 
This formula also defines unitary operators in $L^2(\R^n)$
and continuous transformations of the space $\cS'(\R^n)$
of the space of tempered distributions on $\R^n$.

\SS

{\it Adelic case.}
The group $\Heis_n(\A)$ acts on the space $\cB(\A^n)$
by the formula 
\begin{equation}
T(v_+,v_-,\alpha)f(x)=
f(x+v_+)
\exp\bigl\{2\pi i (xv_-^t+ 
\alpha+ \frac12 w_+  v_-^t)\bigr\}
\label{adeli-heis}
.\end{equation}
This formula also defines unitary operators
in $L^2(\A^n)$ and continuous operators in the space
$\cB'(\A_n)$ of adelic distributions.

\SS

{\it $p$-adic case.}
The action of $\Heis_n(\Q_p)$ on $\cB(\Q_p)$ and $\cB'(\Q_p)$
is defined by the same formula.

\SS

{\it Rational case.}
The group $\Heis_n(\Q)$ acts in the space $\cM(\Q^n)$
via the same formula (\ref{adeli-heis}).


{\bf \punct Relations between the standard representations
of $\Heis_n(\cdot)$.}

\begin{proposition}
a) The subgroup $\Heis_n(\Q)\subset \Heis_n(\R)$
preserves the space $\cP(\R^n)$.

b) The canonical map $I_\R :\cM(\Q^n)\to \cP(\R^n)$
commutes with the action of $\Heis_n(\Q)$.

c) The canonical bijection $\cM(\Q^n)\to \cB(\A^n)$
commutes with the action of $\Heis_n(\Q)$.
\end{proposition}

This statement is more-or-less obvious.
It also implies Theorem \ref{th:main}.a.


{\bf\punct Irreducibility.}

\begin{lemma}
\label{l:lemma}
 The representation of $\Heis_n(\Q)$
in $\cM(\Q^n)$ is irreducible. Any operator $A:\cM(\Q^n)\to\cM(\Q^n)$ 
 commuting with the action of $\Heis_n(\Q)$ is a multiplication
by a constant.
\end{lemma}

{\sc Proof.} First, we present an alternative description of the space $\cM(L|K)$, it 
consists of functions fixed with respect to operators
\begin{align}
T_v f(x)=f(x+v),\qquad v\in K
\label{sdvig}
,
\\
S_w f(x)=f(x)\exp(2\pi i x w^t),\qquad w\in L^\lozenge 
\label{exp}
.
\end{align}
The space  $\cM(L|K)$ is point-wise fixed by the group $G(L|M)$ generated
by these operators.

 The space $\cM(L|K)$ is invariant with respect to
the group $D(L|K)$ generated by the  operators
$T_v$, where $v\in L$, and $S_w$, where $w\in K^\lozenge$.

Hence the quotient-group $A(L|M)=D(L|M)/G(L|M)$ acts in $\cM(L|M)$.
In fact, this group is generated by the same operators $T_v$, $S_w$, see
(\ref{sdvig})--(\ref{exp}), but now we consider $v$ as an element
of $L/M$ and $w$ as an element  of $M^\lozenge/L^\lozenge$
(in fact, $A(L|M)$ is a finite Heisenberg group).

Let us show that the representation of $A(L|M)$ in the space $\cM(L|M)$ 
is irreducible.
The subgroup of $A(L|M)$ generated by the operators $S_w$ 
has a simple specter, its eigenvectors are $\delta$-functions
on $L/M$. Hence any invariant subspace is spanned by some collection
of $\delta$-functions. But $T_v$-invariance implies the triviality
of an invariant subspace.

Now the both statements of the Lemma become obvious.
\QED


\section{Weil representation}

\addtocounter{sec}{1}
\setcounter{equation}{0}
\setcounter{punct}{0}
\setcounter{theorem}{0}

On the Weil representation,
see \cite{Wei},  \cite{Ner}, \cite{KV}, \cite{Naz}.

\SS

{\bf\punct Symplectic groups.} Consider a ring
$\K=\R$, $\Q_p$, $\Q$, $\A$, $\Z$, $\Z_p$.
Consider the space $\K^n\oplus\K^n$ equipped with a skew-symmetric
bilinear form with the matrix $\begin{pmatrix}0&1\\-1&0 \end{pmatrix}$.
By $\Sp(2n,\K)$ we denote the group of all the operators
in $\K^n\oplus\K^n$ preserving this form, we write its
elements as block matrices 
$g=\begin{pmatrix} A&B\\C&D \end{pmatrix}$.

An element of the adelic symplectic group
$\Sp(2n,\A)$ also can be considered as a sequence
$(g_2,g_3,g_5,\dots)$, where
$g_p\in\Sp(2n,\Q_p)$, and $g_p\in\Sp(2n,\Z_p)$ for all
the $p$ except finite number.

\SS


{\bf \punct Automorphisms of the Heisenberg groups.}
Let $\K=\R, \Q,\Q_p,\A$.
The symplectic group $\Sp(2n,\K)$ acts on 
the Heisenberg group $\Heis_n(\K)$ by automorphisms
$$
\sigma(g):\,\, \bigl\{v_+\oplus v_-\bigr\}\oplus \alpha
\mapsto \Bigl\{(v_+\oplus v_-)
\begin{pmatrix} A&B\\C&D \end{pmatrix}\Bigr\}
\oplus \alpha,
$$
see (\ref{heis-product}).

\SS

{\bf \punct Real case.}

\begin{theorem} a)  For each $g\in\Sp(2n,\R)$,
 there is a unique up to a factor unitary operator 
$\We(g):L^2(\R^n)\to L^2(\R^n)$
such that for each $h\in\Heis_n$
\begin{equation}
T(\sigma(h))=\We(g)^{-1} T(h) \We(g)
\label{weil-def}
.
\end{equation}

b) For each $g_1$, $g_2\in \Sp(2n,\R)$,
$$
\We(g_1)\We(g_2)=c(g_1,g_2) \We(g_1g_2)
,$$ 
where $c(g_1,g_2)\in \C$. Moreover, there is a choice of
$\We(g)$, such that $c(g_1,g_2)=\pm 1$ for all $g_1$, $g_2$.
\end{theorem}

Thus $\We(\cdot)$ is a projective representation of $\Sp(2n,\R)$.
It is named the {\it Weil representation.}

It is easy to write the operators $\We(g)$ for some special
matrices $g$, 
\begin{align}
\We\begin{pmatrix}A&0\\0&A^{t-1}\end{pmatrix} f(x)&=
|\det(A)|^{-1/2}  f(xA^{t-1})
\label{weil-r-1}
,\\
\We\begin{pmatrix}  0&1\\-1&0\end{pmatrix} f(x)
&=(2\pi)^{-n/2}
\int_{\R^n} f(y)\exp\{ixy^t\}\,dy
\label{weil-r-2}
,\\
\We\begin{pmatrix}1&B\\0&1\end{pmatrix} f(x)
&=
\exp\{\frac i2 xBx^t\}f(x)
\label{weil-r-3}
,\end{align}
where the matric $B$ is symmetric, $B=B^t$. 

Since these elements generate the whole group
$\Sp(2n,\R)$, our formulae allow to obtain $\We(g)$ for an
arbitrary $g\in\Sp(2n,\R)$.

\begin{theorem} 
\label{th:weil-r}
The space $\cP(\R^{2n})$ is invariant
with respect to the action of the group $\Sp(2n,\Q)$.
\end{theorem}

{\sc Proof.} Obviously, $\cP(\R^n)$ is invariant with respect
to operators (\ref{weil-r-1}), (\ref{weil-r-3}) with rational
matrices $A$, $B$.

By the Poisson summation formula, $\cP(\R^n)$ is invariant with
respect to the Fourier transform (\ref{weil-r-2}).

It can be readily checked that the group $\Sp(2n,\R)$ is generated
by elements of these 3 types, and this finishes the proof.
\hfill $\square$.


\SS

{\bf \punct $p$-adic Weil representation.} For the group 
$\Sp(2n,\Q_p)$, the literal analog
 of Theorem \ref{th:weil-r}  is valid.
In this case the operators $\We(g)$ are unitary in $L^2(\Q_p^n)$
and preserve the Bruhat space $\cB(\Q_p^n)$.

Analogs of formulae (\ref{weil-r-1})--(\ref{weil-r-2})
also can be easily written, 
\begin{align}
\We\begin{pmatrix}A&0\\0&A^{t-1}\end{pmatrix} f(x)&=
 |\det(A)|^{-1/2}  f(xA^{t-1})
\label{weil-q-1}
,\\
\We\begin{pmatrix}  0&1\\-1&0\end{pmatrix} f(x)
&=
\int_{\R^n} f(y)\exp\{2\pi ixy^t\}\,dy
\label{weil-q-2}
,\\
\We\begin{pmatrix}1&B\\0&1\end{pmatrix} f(x)&=
\exp\{\pi i\, xBx^t\}f(x)
\label{weil-q-3}
.\end{align}

{\sc Remark.} After an appropriate normalization
of operators $\We(g)$, we can obtain
\begin{equation}
W(g)\cI_{\Z_p^n} =\cI_{\Z_p^n},\qquad
\text{where $g\in \Sp(2n,\Z_p)$ }
\label{compact-sp}  
,\end{equation} 
see also Proposition \ref{pr:gamma-12}


{\bf\punct Adelic Weil representation.}
We have
$$
L^2(\A^n)=\bigotimes_p \Bigl(L^2(\Q_p^n), \cI_{\Z_p^n} \Bigr)
,\qquad 
\cB(\A^n)=\bigotimes_p \Bigl(L^2(\Q_p^n), \cI_{\Z_p^n} \Bigr)
,$$
in the first case we have a tensor product in the category
of Hilbert spaces, in the second case we have a tensor product
in the category of abstract linear spaces.

\SS

{\sc Remark.} To define a tensor products
of an infinite family of spaces $V_j$, we need in a distingueshed
unit vector $e_j$ in each space, the tensor product space
$\bigotimes V_j$ is spanned by products 
$v_1\otimes v_2\otimes\dots$, where $v_j=e_j$ for all
$j$ except a finite set.\QED

\SS   

The Weil representation of $\Sp(2n,\A)$ is defined
as 
$
W(g)=\bigotimes W(g^{(p)}).
$
These operators are unitary in $L^2(\A^n)$ and preserve
the dense subspace $\cB(\A^n)$. For almost all $\cI_{\Z_p^n}$,
we have $\We(g^{(p)}) \cI_{\Z_p^n} =\cI_{\Z_p^n}$ and this allows
to define tensor products of operators.

\SS


{\bf \punct Proof of Theorem \ref{th:main}.b.}
Transfer the representations of $\Sp(2n,\Q)$
from the spaces $\cB(\A^n)$, $\cM(\R^n)$ to the space
$\cM(\Q^n)$. We obtain two representations of $\Sp(2n,\Q)$
in $\cM(\Q^n)$, say $\We_1(g)$, $\We_2(g)$. These operators
satisfy the commutation relations
$$
T(\sigma(h))=\We_1(g)^{-1} T(h) \We_1(g),\qquad
T(\sigma(h))=\We_2(g)^{-1} T(h) \We_2(g)
.$$
Hence $\We_1(g)^{-1}\We_2(g)$ commutes with $T(h)$.
By Lemma \ref{l:lemma}, 
$\We_2(g)=\lambda(g)\We_1(g)$, where $\lambda\in\C$.
\QED


\section{Addendum. Constructions of modular forms}

\addtocounter{sec}{1}
\setcounter{equation}{0}
\setcounter{punct}{0}
\setcounter{theorem}{0}

Here we explain the standard construction of modular
forms from theta-functions and Howe duality, see
\cite{Sch}, \cite{Igu}, \cite{Mum}, \cite{LV}.

\SS


{\bf\punct Congruence subgroups.} Consider the group
$\Sp(2n,\Z)$ of symplectic matrices
$g=\begin{pmatrix} A&B\\C&D\end{pmatrix}$ with integer elements.
For a positive integer $N$,
 denote by $\Gamma_N$ the {\it principal congruence-subgroup}
consisting of matrices $g\in\Sp(2n,\Z)$ such that $N$
divides all the  matrix
elements of $g-1$.
A {\it congruence subgroup} in $\Sp(2n,\Z)$ 
is any subgroup including a principal congruence-subgroup.

\SS

For the following statement, see, for instance, \cite{Ven}

\begin{theorem}
\label{th:generators}
The subgroup in $U_l\subset\Sp(2n,\Z)$
generated by matrices
\begin{equation}
\begin{pmatrix}
1+l\alpha&0\\0&(1+l\alpha)^{t-1}
\end{pmatrix},
\begin{pmatrix}1&l\beta\\0&1\end{pmatrix},
\begin{pmatrix}1&0\\l\gamma&1\end{pmatrix}
\label{generators-spnz}
,\end{equation}
where $\alpha$, $\beta$, $\gamma$, 
$(1+l\delta)^{-1}$ are integer matrices,
is a congruence subgroup
\end{theorem}


{\bf\punct The subgroup $\Gamma_{1,2}$.}
The denote by $\Gamma_{1,2}$ the subgroup
of $\Sp(2n,\Z)$ consisting of 
$g=\begin{pmatrix}A&B\\C&D\end{pmatrix}$
such that the matrices $A^tC$ and $B^tD$ have 
even elements on the diagonals.
For the following theorem, see \cite{Mum}).

\begin{theorem}
The group $\Gamma_{1,2}$ is generated by
matrices
$$
\begin{pmatrix} 
A&0\\0&A^{t-1}
\end{pmatrix},\quad
\begin{pmatrix}
1&B\\0&1
\end{pmatrix},
\quad
\begin{pmatrix}
1&0\\C&1
\end{pmatrix}
,$$
 where the matrices 
$B$, $C$ have even diagonals.
\end{theorem}


Denote 
$$
\Delta(x)=\sum_{k_1,\dots,k_n}
 \prod_j\delta(x_j-\sqrt{2\pi}k_j) 
$$

\begin{proposition}
\label{pr:gamma-12}
The restriction of the Weil representation of $\Sp(2n,\R)$
to $\Gamma_{1,2}$ is a linear representation.
Moreover, we can normalize the operators $\We(g)$, $g\in\Gamma_{1,2}$,
such that 
\begin{equation}
\We(g)\Delta =\Delta
\label{we-delta}
\end{equation}
\end{proposition}

{\sc Proof.} First, $\Delta$ is an eigenvector for
operators $\We(g)$, $g\in\Gamma_{1,2}$. It is easy to verify
this for generators of $\Gamma_{1,2}$, and hence this is valid
for all $g$. Now we can choose the normalization
(\ref{we-delta}). Now $\We(g)$ 
became a linear representation of $\Gamma_{1,2}$.
\QED


{\bf \punct Congruence subgroups
 and the space $\cP(\R^n)$.}

\begin{theorem}
\label{th:modular-origin}
The stabilizer of each element of $\cP(\R^n)$
in the group $\Gamma_{1,2}$ is 
a congruence subgroup.
\end{theorem}

{\sc Proof.} It is easy to verify (see Theorem \ref{th:generators})
that the subgroup $U_{2N^2}$ fix all the vectors of
$\cP(N^{-1}\Z^n|N Z^n)$.\QED

\SS


{\bf \punct Modular forms of the weight $1/2$.}
Denote by $W_n$ the Siegel upper half-plane, i.e., the set of
$n\times n$ complex matrices satisfying the condition
$\frac 1{2i}(z-z^*)>0$. The group $\Sp(2n,\R)$ acts
in the space of holomorphic functions on $W_n$
by the following operators
\begin{equation}
T_{1/2}\begin{pmatrix} A&B\\C&D\end{pmatrix}
f(x)=f((A+zC)^{-1}(B+zD))\det(A+zC)^{-1/2}
\label{1/2}
\end{equation}

Consider the operator
$$
J\chi(z)= \bigl\{\exp\bigl(\frac 12 xzx^t \bigr) ,\chi \bigl\}
$$
from $\cS'(\R^n)$ to our space of holomorphic functions.
It is easy to verify, that this operator intertwines the
Weil representation and the representation $T_{1/2}$.

By Proposition \ref{pr:gamma-12}, for $g\in\Gamma_{1,2}$,
we can normalize the operators (\ref{1/2})
$$T_{1/2}'(g)= \lambda(g) T_{1/2}(g),\qquad \lambda(g)\in\C$$
and obtain a linear representation of $\Gamma_{1,2}$
(in fact, $\lambda(g)$ ranges in 8-th roots of 1).

\begin{proposition}
Let $\chi\in\cP(\R^n)$ be a Poisson distribtion,
$\Phi=J\chi$. There is a congruence subgroup 
$\Gamma\subset\Gamma_{12}$ such that 
$$
T'_{1/2}(g)\Phi=\Phi,\qquad \text{where $g\in\Gamma$}
$$
 \end{proposition} 

\SS

In fact, Theorem \ref{th:modular-origin} provides lot of 
possibilities to produce modular forms.
For instance, consider some embedding $I:\SL(2,\R)\to \Sp(2n,\R)$
such that $i(\SL(2,\Q))\subset\Sp(2n,\Q)$.
Assume that the restriction of the Weil represntation
to $\SL(2,\R)$ contains a subrepresentation $V$ of a discrete series%
\footnote{On representations of $\SL(2,\R)$,
see, for instance \cite{GGP}.}.
Then we can consider projection of the space $\cP(\R^n)$ to
$V$.

\medskip

{\sf Math.Physics group, 
 Institute of Theoretical and Experimental Physics, %
\linebreak
B.Cheremushkinskaya, 25, Moscow 117 259, Russia


\&
Math.Dept,
University of Vienna,
Nordbergstrasse, 15, Vienna }

\tt neretin@mccme.ru

\end{document}